\documentclass[intlimits,twoside,a4paper]{article}

\usepackage{amsmath,amssymb}
\usepackage{graphicx}
\usepackage{hhline}

\usepackage[T2A]{fontenc}
\usepackage[cp1251]{inputenc}

\usepackage[eqsecnum]{cmpj2}

\issue{2016}{19}{4}{43704}
\doinumber{10.5488/CMP.19.43704}

\title[Dynamic conductivity of one-dimensional ion conductors]%
{Dynamic conductivity of one-dimensional ion conductors.
Im\-pe\-dan\-ce, Nyquist diagrams}
\author{I.V. Stasyuk, R.Ya. Stetsiv}
\address{Institute for Condensed Matter Physics
of the National Academy of Sciences of Ukraine,\\
1 Svientsitskii St., 79011 Lviv, Ukraine}

\authorcopyright{I.V. Stasyuk, R.Ya. Stetsiv, 2016}
\date{Received September 23, 2016}

\begin{document}

\maketitle

\begin{abstract}
Dynamic conductivity of the one-dimensional ion conductor is
investigated at different values of the interaction constant
between particles and the modulating field. The consideration is
based on the hard-core boson lattice model. Calculations are
performed for finite one-dimensional cluster using the exact
diagonalization method. Frequency dependence of the dynamic
conductivity and behaviour of its static component (Drude weight)
in the charge-density-wave (CDW) and superfluid (SF) phases are
studied. Frequency dispersion of impedance and loss tangent is
calculated; the Nyquist diagrams are built and analyzed.
\pacs 75.10.Pq, 03.75.Lm, 66.30.Dn
\end{abstract}

\section{Introduction}
Transport properties of one-dimensional systems are of great
interest over many years
\cite{Mahan,Maldague,Shastry,Castella,Giamarchi1,Giamarchi2,Kuhner,Heidrich,Mukerjee}.
One-dimensional objects have special features, and, in addition,
the relevant models are investigated much easier than for higher
dimensionalities and in many cases permit to obtain  exact
solutions. The resulting solutions frequently provide the understanding
of conduction for 2d and 3d  systems, but,  at the same time,
there exist purely one-dimensional complex problems. The systems
with ionic conductivity are of special interest. Attention to
these systems is paid due to ever-increasing possibilities of
their practical applications  as solid electrolytes in capacitors and
batteries, in membranes of fuel cells \cite{Padma}, in
electronics, in controlling and signalling devices for special purposes.
Therefore, new compounds with high ionic conductivity were recently
synthesized in order to find materials stable against
chemical and mechanical action and possessing other specific
properties. Just recently a new superionic crystal
Li$_{10}$GeP$_{2}$S$_{12}$, the conductivity of which reaches 12~m\textohm$^{-1}$\,cm$^{-1}$ at room temperature and 0.41~m\textohm$^{-1}$\,cm$^{-1}$
at $-30^\circ$C, was synthesized \cite{Kanno}. The conductivity of
ionic conductors is especially high when a number of ions is much
smaller than the number of positions in a lattice, i.e., when there
are vacancies. Therefore, a lot of free positions facilitate the
ion hopping probability from one position to another. Charge
transfer process in some superionics occurs along the chain
(one-dimensional) structures. The proton conductor
LiN$_{2}$H$_{5}$SO$_{4}$ \cite{Pietraszko}, some superionic
(superprotonic) conductors, in particular CsHSO$_{4}$
\cite{Lahajnar}, coordination polymers such as iron oxalate dihydrate
Fe(C$_{2}$O$_{4}$)$\,\cdot\,$2H$_{2}$O, nanotubes  \cite{Yamada}, etc. are the
examples. One-dimensional system with Josephson junctions (one
contact in width and several hundred tunnel contacts in length)
are created \cite{Chow}.  In most cases, the quantum systems
are studied at low temperatures, when the transport properties of
a system are determined mainly by their ground state.
Conductivity at high temperatures is less studied. In such a case
one  cannot restrict oneself to strictly defined elementary
excitations (see for example \cite{Mukerjee}).

Lattice models are widely used for a theoretical description of
ion and proton transport at the microscopic level. As a rule, the
Bose-Hubbard model is used here at arbitrary occupation of local
particle positions (see review \cite{Bloch}). There are also used
the models based on Fermi statistics
\cite{Salejda1,Stasyuk1,Krasnogol,Stasyuk2,Stasyuk4} or on the
``mixed'' Pauli statistics
\cite{Mahan,Stasyuk5,Stasyuk6,Stetsiv,Micnas1,Micnas2,Rigol,Hen1,Hen2,Batrouni1,Batrouni2,Bernardet1,Bernardet2,Schmid,GUO},
where particles are of Bose nature, but they obey the Fermi
rule as well. The lattice model of Pauli particles is similar to the
Bose-Hubbard model in the hard-core approximation (provided that
the occupation numbers are restricted, $n_{i}$ = $0,1$). When
taking into account only the on-site interaction ($U$) of particles
in the one-dimensional Bose-Hubbard model,   the Mott insulator (MI)
phase with the integer particle density was received. At
intermediate concentrations, the superfluid (SF) type state [the
phase with the infinitely large (divergent) boson correlation
length and without the order parameter (see for example
\cite{Kuhner})] appears at $T = 0$. Inclusion of the
near-neighbors interaction ($V$) leads to charge-density-wave
phase (CDW) with half-filling of ionic sites on the average. The
Pauli (hard-core boson) lattice model also enables one to describe the
transitions (which are the true ones only at $T = 0$) between
these phases, including the emergence of the SF-type state that
can appear even in the absence of a direct interaction between
particles \cite{Stasyuk5,Stasyuk6,Micnas1}.

 In our previous work \cite{Stetsiv} in the hard-core boson approximation, the one-particle spectral density was
calculated. The exact diagonalization method for finite
one-dimensional lattice model with periodic boundary conditions
was used.  The conditions of existence of various phases of
a system, depending on the values of interaction between
particles $V$ and the modulating field strength $A$, were established
by analyzing the character of spectral density; the phase diagrams
were built \cite{Stetsiv}. The results agree with the known ones
from the literature. In particular, the diagram of state for the
case $V = 0$ coincides with the exact diagram obtained
analytically for the one-dimensional case, see for example
\cite{Hen2} (the exact analytical solution can be obtained in this
case by applying the Jordan-Wigner transformation, which makes it
possible to pass from the Hamiltonian of hard-core bosons to the
Hamiltonian of noninteracting spinless fermions). It was shown in
\cite{Stetsiv} that at $T = 0$, the repulsive short-range
interaction between particles ($V > 0$) results in the emergence
of a gap in the energy spectrum in the limit of half-filling ($n
= 1/2$). A true CDW phase is realized here only at zero
temperature. The gap in the CDW-phase spectrum grows if the
magnitude of either the short-range interaction $V$ or the
modulating field $A$ (which can be associated with an internal
field that appears owing to the long-range interaction) increases.
At $T \neq 0$, the gap gradually disappears with the temperature
rise; the interphase boundaries on the ($T,\,\mu$) plane become
smeared, and the corresponding phase transitions in this case
reduce to the crossover transformations, i.e., they are not
genuine phase transitions.

This work is devoted to the calculation of static and dynamic
conductivity of one-dimensional ion conductor. The attention
was mostly focused on the study of the static conductivity
$\sigma(\omega = 0)$ described by the so-called Drude weight $D$
\cite{Heidrich,Kampf,Scalapino1,Shastry,Stafford,Fye}. The value
of $D$ usually serves as a criterion that determines the state of a
system: superconducting (superfluid), metallic, or dielectric
(insulator). For infinite systems $D=0$ in the case of insulator
(the MI phase or CDW phase) and $D$ is finite in the case of a
conducting state (SF phase). For 1d systems of finite length ($L$),
the Drude weight $D$ goes exponentially to zero at
$L\rightarrow\infty$ in an insulator state (the phase with a gap in
spectrum), but remains finite in this limit in the conducting one
(the gapless phase), \cite{Shastry,Stafford,Fye}. The
investigations of the frequency dependence of the dynamic part of
conductivity $\sigma_{\text{reg}}(\omega)$ in the MI phase show that this quantity
is equal to zero at low frequencies and becomes nonzero starting
from some threshold value of $\omega$ \cite{Maldague,Fye}.
Contrary to that, in the SF phase, the $\sigma_{\text{reg}}(\omega)$
function increases with frequency according to the power law starting
from the zero value at $\omega =0$ \cite{Giamarchi2} and decreases exponentially
at high frequencies. Besides that, the existence of peaks on the
frequency dependence of $\sigma_{\text{reg}}(\omega)$ was noticeable. For 1d
models with finite on-site repulsion $U$, the presence of three
peaks was shown in \cite{Kuhner}; the peaks are located in the
frequency region $\hbar \omega \sim U$. The calculations,
performed in the case of hard-core bosons for MI phase, revealed
one (or two, depending on approximation) peak of
$\sigma_{\text{reg}}(\omega)$ \cite{Lindner}.

Our lattice model includes the ion transfer, the interaction between the
neighboring ions, and the modulating field. By applying the exact
diagonalization technique we determine the energy spectrum and
matrix elements of the current density operator for the finite
one-dimensional model of the ionic conductor. Based on the Kubo
theory \cite{Kubo}, we numerically calculate the static (Drude
weight) and dynamic (frequency dependent) parts of conductivity.
The main attention is paid to the investigation of the ion
conductivity in the SF and  CDW phases and to the study of
the effect of the short-range interaction between particles as well
as the modulating field (that effectively appears due to
long-range interaction or due to the two-sublattice
structure). Restricting ourselves to the $T=0$ case, we analyze
the differences between the Drude weight values and frequency
dispersion of conductivity in  CDW and SF phases. In addition,
the plots of impedance are presented and Nyquist diagrams are
built. In our consideration, we restrict ourselves in this work to
the case $T=0$.

\section{Hamiltonian and its transformation}

We consider the one-dimensional ion conductor as the chain of
heavy immobile ionic groups and light ions that move along this
chain occupying certain positions between the mentioned groups.
The subsystem of light ions is described with the following
Hamiltonian
\begin{eqnarray}
\hat{H} &=& t \sum\limits_{i} (c_{i}^{+}c_{i+1} + c_{i+1}^{+} c_{i}) + V \sum_{i}n_{i}n_{i+1} - \mu \sum_{i} n_i
+ A \sum_{i} (-1)^{i}n_i \label{ham1}.
\end{eqnarray}
Here, the operators of creation and annihilation of particles
($c^+_{i,\alpha}$ and $c_{i,\alpha}$)  obey the Pauli statistics.
The model (\ref{ham1}) is equivalent to the hard-core boson model.
The model takes into account the nearest-neigh\-bour ion transfer
(with hopping parameter $t$), interaction between ions that occupy
nearest-neighbouring positions (with corresponding parameter $V >
0$) and modulating field (parameter $A$). The system is divided
into two sublattices under the effect of the $A$ field, which
simulates the long-range interactions between the particles, which
contributes to the modulation of the spatial distribution of light
ions in the so-called ordered phase (the existence of such phases
at low temperatures is a characteristic feature of superionic
conductors).

In order to calculate the energy spectrum of the one-dimensional
ionic Pauli conductor we apply the exact diagonalization
technique. For this purpose, let us consider a finite chain with
periodic boundary conditions. For a chain with $N$ sites in the
main region, we introduce the many-particle states
\begin{equation}
\mid n_{1}\,, n_{2}\,, \ldots, n_{N} \rangle .
\end{equation}
The Hamiltonian matrix on the basis of these states is the matrix
of the order $2^N \times 2^N$. This matrix is numerically diagonalized
\cite{Stetsiv}. Such an operation corresponds to the
transformation
\begin{eqnarray}
U^{-1} H U = \widetilde{H} = \sum \limits_p E_p \widetilde{X}^{pp},
\end{eqnarray}
where $E_p$ are eigenvalues of the Hamiltonian,
$\widetilde{X}^{pp}$ are Hubbard operators. The same
transformation is applied to the operators of particle creation
and annihilation at the $i$-th chain site
\begin{eqnarray}
U^{-1} c_{i} U = \sum \limits_{p,q} A^{i}_{pq} \widetilde{X}^{pq}, \qquad \quad U^{-1} c^+_{i} U = \sum
\limits_{r,s} A^{i*}_{rs} \widetilde{X}^{rs}.
\end{eqnarray}
Herein below we shall express the current density and conductivity
operators applying this representation. It should be marked that
we used it in \cite{Stetsiv} to calculate the single-particle
spectral density
$\rho(\omega) = -2\Im \langle\langle c|c^+ \rangle\rangle$
for the
model (\ref{ham1}) and analyze its form in different phases of the
system.

\section{Dynamic conductivity. General relations}

The dynamic conductivity of one-dimensional ion conductor can be
calculated starting from the  Kubo formula \cite{Kubo}
\begin{eqnarray}
\sigma(\omega,T) =  \int\limits_{-\infty}^0 \rd t \exp [\ri(\omega - \ri\varepsilon)t] \int\limits_0^{\beta} \rd\lambda
\langle \hat I(t-\ri\,\hbar\lambda)\hat I(0)\rangle, \label{sigKubo}
\end{eqnarray}
here, $\hat I$ is the current operator. We use the following
expression for  current density  operator $\hat j$ (see for
example \cite{Mahan,Kuhner,Scalapino})
\begin{eqnarray}
\hat j(0) = \frac{\ri}{\hbar} t q a \frac{1}{V_{0}} \sum\limits_i \left(c^+_{i} c_{i + 1} - c^+_{i + 1} c_{i}\right),
\end{eqnarray}
here, $q$  is the ion charge, $a$  is lattice constant, $\hat I =
\hat jS$, where $S$ is the conductor cross-section, $V_{0} = SNa$.
The current operator written on the transformed basis is of the form
\begin{eqnarray} \hat I(0) = \frac{\ri}{\hbar} t q \frac{1}{N} \sum\limits_{i=1}^N   \sum\limits_{k,l} \sum \limits_{m} \left[A^{i*}_{km} A^{i + 1}_{kl} - A^{(i +1)*}_{km} A^{i}_{kl}\right] \widetilde{X}^{ml},
\end{eqnarray}
\begin{eqnarray}
I(t) = \re^{\frac{\ri}{\hbar}Ht}I(0)\re^{-\frac{\ri}{\hbar}Ht}.
\end{eqnarray}

According to Kubo formula (\ref{sigKubo}), we obtained the
following expressions for the ion conductivity
\begin{align}
 \sigma(\omega) =  \ri \left(\frac{t q}{\hbar N}\right)^2 \sum\limits_{i=1}^N
\sum\limits_{j=1}^N \sum\limits_{k,l} \sum \limits_{m,n} &\left[ A^{i*}_{km} A^{i + 1}_{kl} - A^{(i +1)*}_{km}
A^{i}_{kl}\right] \left[ A^{j*}_{nl} A^{j + 1}_{nm} - A^{(j +1)*}_{nl} A^{j}_{nm}\right] \nonumber
\\
&\times\frac{1}{Z} \frac{1}{E_{l} - E_{m}}
\frac{\re^{-\beta E_m} - \re^{-\beta E_l}}{\omega - \frac{1}{\hbar} (E_{m} - E_{l}) + \ri\varepsilon}\,. \label{sigma}
\end{align}
The real part of the dynamic conductivity exhibits a  discrete
structure that  consists of many $\delta$-peaks in the case of
finite size of a cluster. If the chain size (the number of sites
$N$) increases, the $\delta$-peaks are located more densely and,
at $N = \infty$, form a band structure. In our calculations, we
confined ourselves to the case $N = 10$.  The small parameter
$\Delta$ was also introduced to broaden the $\delta$-peaks
according to Lorentz distribution
\begin{equation}
\delta (\omega) \rightarrow \frac{1}{\pi} \frac{\Delta}{\omega^2+\Delta^2}\,.
\end{equation}
In what follows, we relate all energy parameters, including
$\hbar \omega$, to the hopping constant $t$, which is taken as
the energy unit. For convenience, we use the notation  $\mu' = \mu
- V$.

The expression for the static conductivity can be obtained when in
the formula (\ref{sigma}) we put $E_{m} = E_{l}$, taking into
account only the contribution of degenerate states.

Real part of the conductivity will possess in this case the
component proportional to $\delta(\omega)$:
\begin{align}
 \Re \sigma(\omega \rightarrow 0) =  \frac{1}{Z}  {\pi}  \beta  \left(\frac{t q}{\hbar N}\right)^2 \sum\limits_{i=1}^N
\sum\limits_{j=1}^N \sum\limits_{k,l} \sum \limits_{m,n} &\left[ A^{i*}_{km} A^{i + 1}_{kl} - A^{(i +1)*}_{km}
A^{i}_{kl}\right] \left[ A^{j*}_{nl} A^{j + 1}_{nm} - A^{(j +1)*}_{nl} A^{j}_{nm}\right]  \nonumber
\\
&\times\re^{-\beta E_m} \cdot \delta(\omega) \equiv D
\cdot \delta(\omega). \label{sigmaD}
\end{align}
Here, the summing over indices $l, m$ is performed only for $E_{m}
= E_{l}$. As can be seen from the formula (\ref{sigmaD}), Drude
weight $D$ is different from zero at $T \neq 0$  if the degenerate states are
present (at $T = 0$  $D \neq 0$ when the
ground state is degenerated).

\section{Drude weight}

We analyzed the energy spectrum obtained for different phases of
our model. In all cases, the ground state is nondegenerate (that
can be due to a finite size of the chain structure). Therefore, at $T =
0$ according to formula (\ref{sigmaD}), the static conductivity is
equal to zero, $\sigma(\omega = 0, T = 0) = 0$. The expression
(\ref{sigmaD}) is similar to the one for Drude weight $D^{\,\text{I}}$
obtained in the work \cite{Heidrich}. However, as indicated in
\cite{Heidrich}, the expression for $D^{\,\text{I}}$ does not describe the
temperature dependence of the Drude weight at low temperatures.
Based on the theory of linear response, an
alternative expression for Drude weight was obtained in \cite{Heidrich}
\begin{eqnarray}
D^{\,\text{II}}(N,T) = \frac{\pi}{N} \Bigg[\langle - \hat T \rangle - \frac{2}{Z} \sum  \limits_{m,n \atop E_n \neq E_m }
\re^{-\beta E_n} \frac{\mid \langle m \mid j \mid n \rangle \mid^2}{E_n - E_m} \Bigg].
\label{sigmaDII}
\end{eqnarray}
At the summing over indices $n$, $m$, only the states with $E_{m}
\neq E_{n}$ are taken into account. $\hat T$ is the operator of
kinetic energy
\begin{eqnarray}
\hat T = t \sum\limits_{i} \left(c_{i}^{+}c_{i+1} + c_{i+1}^{+} c_{i}\right). \label{T}
\end{eqnarray}

In the derivation of relation (\ref{sigmaDII}), the $f$-sum rule
(see for example \cite{Shastry,Maldague})
\begin{eqnarray}
\int\limits_{-\infty}^{\infty} \Re \sigma(\omega) \rd\omega = \frac{\pi e^2}{d \hbar^2 L^d} \langle - \hat T
\rangle \label{Shastr}
\end{eqnarray}
was used. Here, $d$ is the dimensionality of the system, $L$ is
the linear dimension. In a number of other works the coefficient at
the average value of kinetic energy is often different. It is due
to adoption of a system of units in which the constants $\hbar$, and
$c$ are taken equal to unity ($\hbar = c = 1$) and the lattice
constant is taken as $a = 1$ (see for example \cite{Scalapino}).

Both quantities of $D^{\,\text{I}}$ and $D^{\,\text{II}}$ coincide in thermodynamic
limit
\begin{eqnarray}
 D(T) = \lim\limits_{N \rightarrow \infty} D^{\,\text{I}}(N,T) = \lim\limits_{N \rightarrow \infty} D^{\,\text{II}}(N,T),
\end{eqnarray}
but they are non-equivalent for finite system \cite{Heidrich}. In
the paper \cite{Heidrich} it was shown that the difference between
$D^{\,\text{I}}$ and $D^{\,\text{II}}$ is negligibly small  for finite system at
high temperatures. At low temperatures, only $D^{\,\text{II}}$ exhibits the
correct temperature dependence. Nevertheless, there are also
some problems. In particular, when approximating $T \rightarrow 0$,
the  value $D^{\,\text{II}}$ for finite systems is often negative
\cite{Heidrich,Scalapino}.

As a whole, the real part of the conductivity can be written in
the form
\begin{eqnarray}
 \Re \sigma(\omega)  = D \cdot \delta(\omega) + \Re \sigma_{\text{reg}}(\omega), \label{reg}
\end{eqnarray}
where besides the Drude term, the regular part $\Re
\sigma_{\text{reg}}(\omega)$ is present.

Integrating the relation (\ref{reg}) over the frequency and using
the formula (\ref{Shastr}), one can write the expression for the
Drude weight in the form equivalent to $D^{\,\text{II}}$ (see for example
\cite{Kuhner})
\begin{eqnarray}
 D = \frac{\pi q^2}{ \hbar^2 N} \langle - \hat T \rangle - \int\limits_{-\infty}^{\infty} \Re \sigma_{\text{reg}}(\omega) \rd\omega. \label{DDD}
\end{eqnarray}

\begin{figure}[!b]
\centerline{\includegraphics[width=0.5\columnwidth]{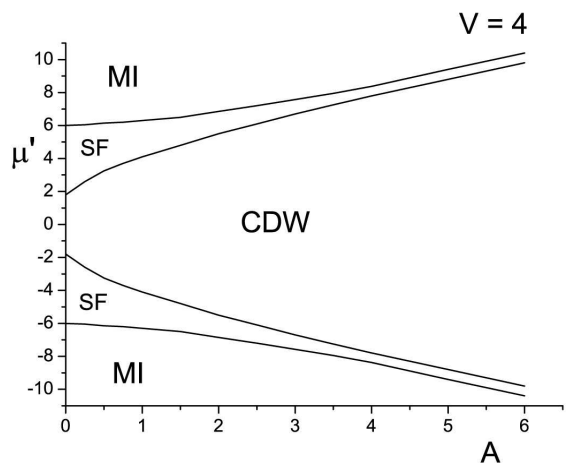}}
\caption{State diagram for a one-dimensional model
of ionic conductor at $T = 0$; $V = 4$, $t = 1$ \cite{Stetsiv}.} \label{fig1}
\end{figure}

We performed the numerical calculation of the Drude weight for our
model of one-dimensional ionic conductor  in different phases
based on the formula (\ref{DDD}). For  the Mott insulator (MI) phase
(see the phase diagrams \cite{Stetsiv} and, in particular,
the diagram in figure~1) we have got $\langle \hat T \rangle = 0$
and $\sigma(\omega) = 0$ at $T = 0$; therefore, $D = 0$, exactly as one
could expect. At half-filling in CDW phase, the system is also an
insulator, so it is expected that in this phase the Drude weight
should be also absent (static conductivity must be equal to zero,
$D = 0$); see for example \cite{Heidrich,Scalapino}. However, as
follows from our calculations, the Drude weight remains finite ($D
> 0$) in the case of half-filled CDW phase, but is significantly
smaller than in the superfluid (SF) phase. Apparently, this is due to
the finite and small size of our one-dimensional cluster ($N =
10$). In particular, at $V = 4$, $\mu' = 0$, $A =5$ for CDW phase
we have obtained $ \frac{1}{N} \langle - \hat T \rangle = 0.139$,
$\int\Re \sigma_{\text{reg}}(\omega) \rd\omega = 0.043$ and
accordingly $D = 0.096$. The gap in the CDW phase increases with
the modulating field $A$ growth. This leads to the reduction of
the calculated Drude weight. We also observe the vanishing of Drude
weight for large values of $A$.

For SF phase, if $ V = 4$, $\mu'= -5$ (see figure~\ref{fig1}), we obtained $D =
0.501$ $(A = 0)$, and $D = 0.416$ $(A = 1)$. The large magnitudes
of static conductivity $\sigma(0)$ in SF phase are also found
for other values of parameters of the model. More results are
shown in tables \ref{tbl} and \ref{tb2}. Hereinafter, we present the
conductivity in relative units omitting the multiplier $\pi q^2/\hbar^2$ and taking $t = 1$. As it follows from the tables,
the static conductivity $\sigma(0)$ in SF phase is determined
mainly by the mean kinetic energy of ions, while in CDW phase
both terms in $D$ should be equal in modulus and compensate each
other.

\begin{table}[!h]
\vspace{-3mm}
\caption{Static conductivity (Drude weight) of one-dimensional
model of ionic conductor in SF phase ($V=4$, $\mu' = -5$, $t = 1
$) at different  values of the modulating field $A$.} \label{tbl}
\vspace{2ex}
\begin{center}
\begin{tabular}{|c|c|c|c|}
\hline\hline $A$ &  $\frac{1}{N} \langle - \hat T \rangle $ & $\int\Re \sigma_{\text{reg}}(\omega) \rd\omega $ & $D$ \\
\hline\hline
0 & 0.501 & 0.0003 & 0.501 \\
\hline
1 & 0.424 & 0.008 & 0.416 \\
\hline\hline
\end{tabular}
\end{center}
\vspace{-3mm}
\end{table}
\begin{table}[!h]
\caption{Static conductivity (Drude weight) of one-dimensional
model of ionic conductor in SF phase ($A = 0$, $\mu' = -1$, $t =
1$) at different values of interaction constant $V$.} \label{tb2}
\vspace{2ex}
\begin{center}
\begin{tabular}{|c|c|c|c|}
\hline\hline
$V$ &  $\frac{1}{N} \langle - \hat T \rangle $ & $ \int\Re \sigma_{\text{reg}}(\omega) \rd\omega $ & $D$ \\
\hline\hline
0 & 0.524 & 0 & 0.524 \\
\hline
1 & 0.607 & 0.0004 & 0.607 \\
\hline
2 & 0.588 & 0.001 & 0.587 \\
\hline\hline
\end{tabular}
\end{center}
\vspace{-3mm}
\end{table}

In the absence of interaction between the ions ($V = 0$), as well
as the modulating field ($A = 0$), the current operator commutes
with the Hamiltonian and is the integral of motion. Then,
$\sigma(\omega \neq 0) = 0$ and Drude weight is determined only by
the ion kinetic energy. In particular, we have obtained along the
lines $V = 0$, $A = 0$ (see corresponding diagrams
\cite{Stetsiv}):  $ D = \frac{1}{N} \langle - \hat T \rangle =
0.524 $ at $\mu' = -1$,  and $ D = \frac{1}{N} \langle - \hat T
\rangle = 0.380$ at $\mu' = -1.5$. This case corresponds to
SF phase.

\section{Dynamic part of conductivity}

\begin{figure}[!t]
\centerline{\includegraphics[width=0.33\columnwidth]{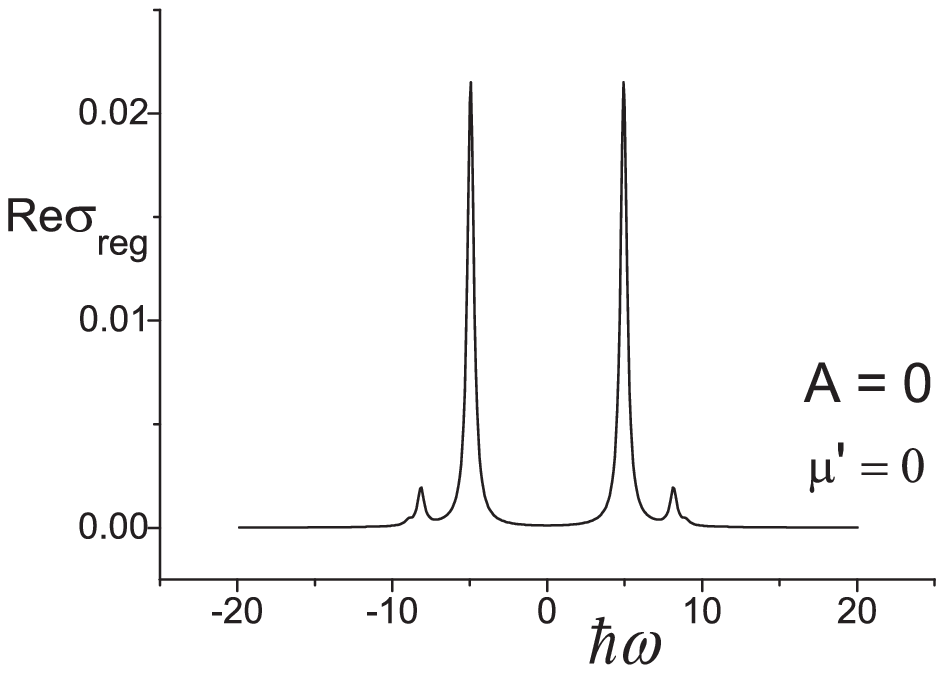}
\includegraphics[width=0.33\columnwidth]{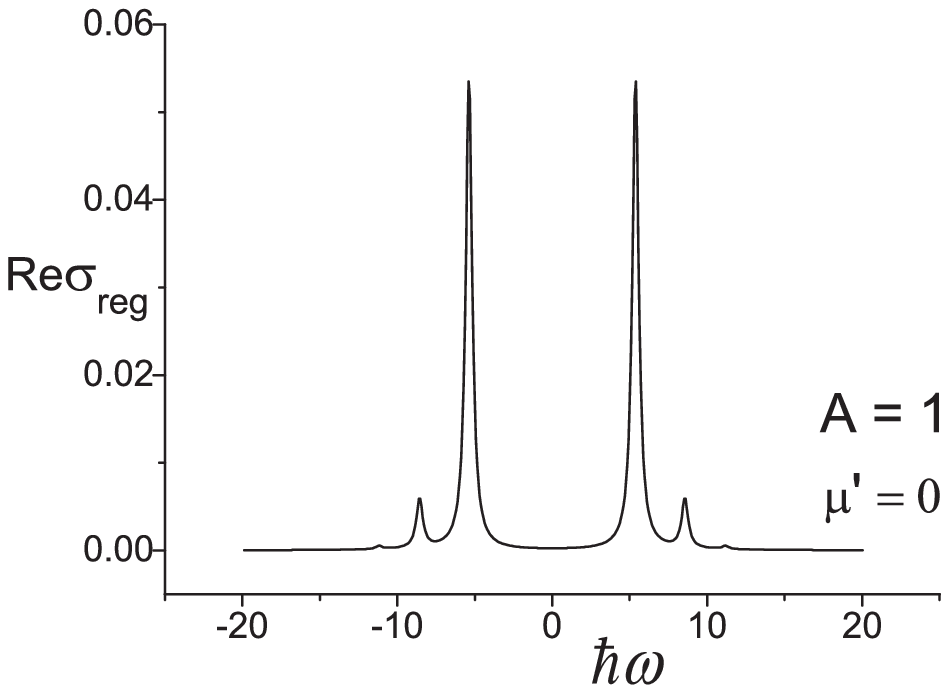}
\includegraphics[width=0.33\columnwidth]{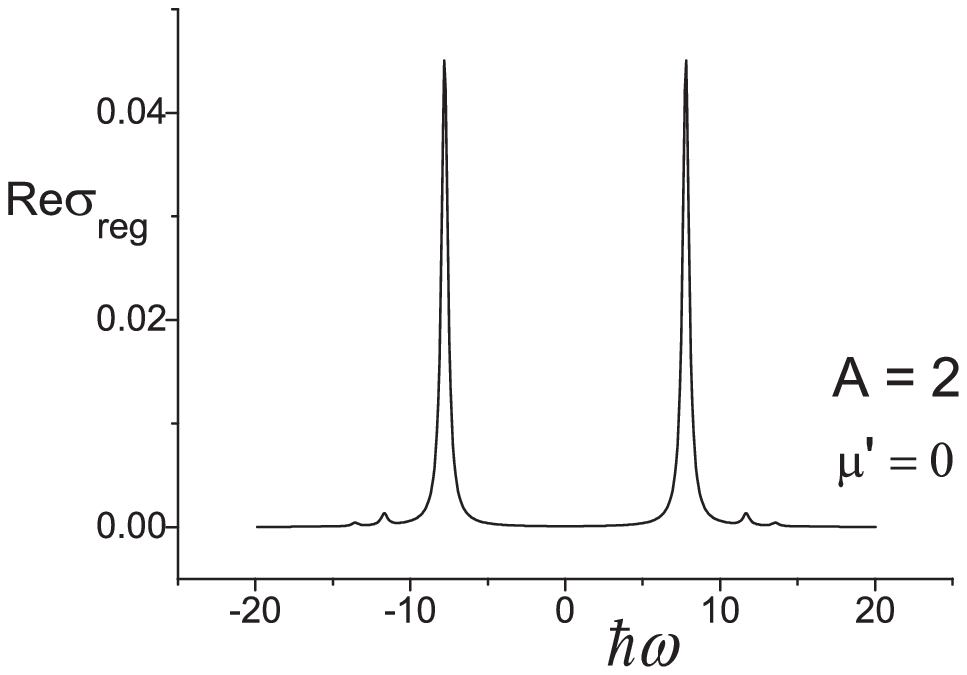}}
\centerline{\includegraphics[width=0.33\columnwidth]{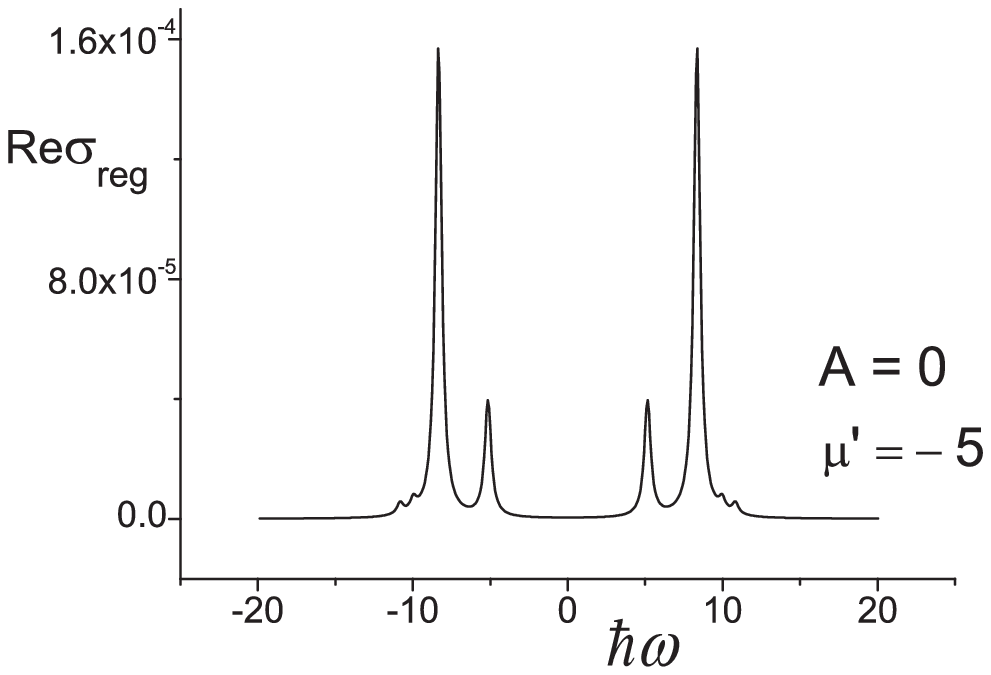}
\includegraphics[width=0.33\columnwidth]{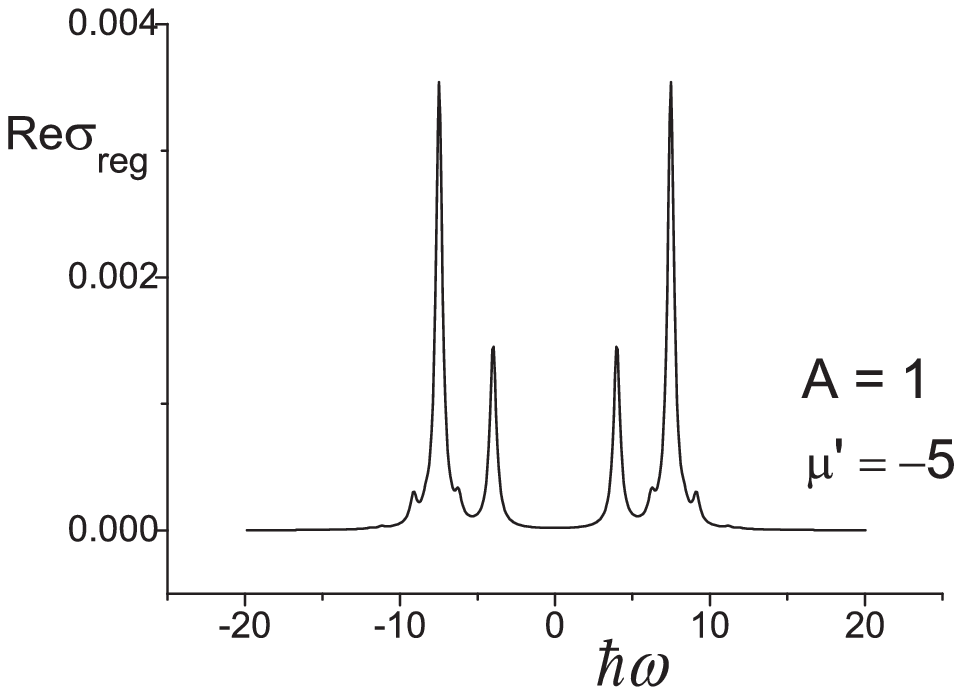}
\includegraphics[width=0.33\columnwidth]{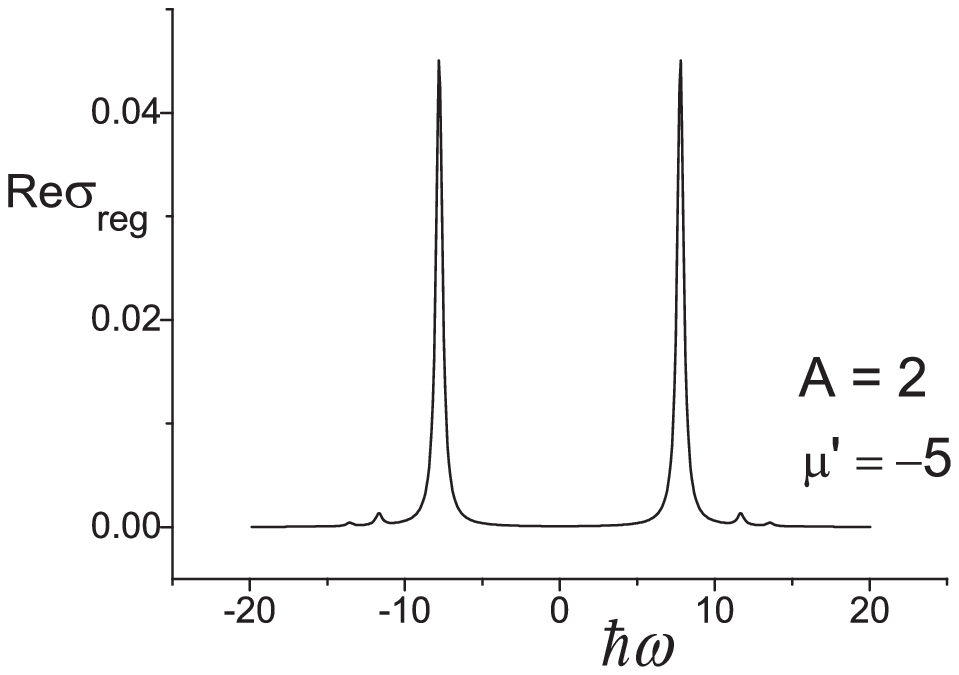}}
\caption{Frequency dependence of the real part of dynamic conductivity of one-dimensional model of the ion
conductor at different values of a modulating field  $A$ in CDW phase ($\mu' = 0$), and in SF phase ($\mu' =
-5$); $N = 10$, $V = 4$, $t = 1$, $\Delta = 0.25$.} \label{fig2}
\end{figure}

\begin{table}[!t]
\caption{The positions of peaks of dynamic conductivity of
one-dimensional model of the ion conductor in CDW phase at
different values of the modulating field  $A$ and fixed
interaction constant $V$: $V=4$ (table on the left) and $V=0$ (table on the right); ($\mu' = 0$, $t = 1 $). Let us mention
that dashes in certain places in the tables mean a  lack of
peaks.}
\label{tb3}
\vspace{2ex}
\begin{center}
\begin{tabular}{|c|c|c|c|c|c|}
\hhline{===~==}
$A$ &  $\hbar \omega_{1}$ & $ \hbar \omega_{2}$ & \hspace{7mm} &  $A$ &  $\hbar \omega_{1}$ \\
\hhline{===~==}
0 & 4.92 & 8.12 &&0 & --\\
\cline{1-3} \cline{5-6}
1 & 5.4 & 8.6 &&1 & 2.36\\
\cline{1-3} \cline{5-6}
2 & 7.8 & 11.72 &&2 & 4.2\\
\cline{1-3} \cline{5-6}
3 & 9.96 & -- &&3 & 6.12\\
\cline{1-3} \cline{5-6}
4 & 11.96 & -- &&4 & 8.12\\
\cline{1-3} \cline{5-6}
5 & 14.04 & -- &&5 & 10.12\\
\hhline{===~==}
\end{tabular}%
\end{center}
\end{table}

\begin{table}[!t]
\caption{The positions of peaks of dynamic conductivity of
one-dimensional model of the ion conductor in SF phase at
different values of the modulating field  $A$ and fixed
interaction constant $V=4$; ($\mu' = -5$, $t = 1 $).} \label{tb4}
\begin{center}
\begin{tabular}{|c|c|c|}
\hhline{===}
$A$ &  $\hbar \omega_{1}$ & $ \hbar \omega_{2}$ \\
\hline\hline
0 & 5.16 & 8.36 \\
\hline
0.5 & 3.64 & 7.0\\
\hline
1 & 3.96 & 7.48 \\
\hline
1.2 & 4.2 & 7.72 \\
\hline
2 & 7.8 & 11.72 \\
\hhline{===}
\end{tabular}
\end{center}
\end{table}

\begin{table}[!t]
\caption{The positions of peaks of dynamic conductivity of
one-dimensional model of the ion conductor in CDW and SF phases
at different values of the interaction constant $V$ and at the absence of a
modulating field ($A = 0$, $t = 1 $): CDW phase at $\mu' = 0$ (the left table) and SF phase at $\mu' = -1$
(the right table).}
\label{tb5}
\vspace{2ex}
\centering
\begin{minipage}{0.26\linewidth}
\begin{center}
\begin{tabular}{|c|c|c|}
\hline\hline
$V$ &  $\hbar \omega_{1}$ & $ \hbar \omega_{2}$ \\
\hline\hline
0 & -- & -- \\
\hline
1 & 4.36& 6.04\\
\hline
2 & 4.36 & 6.68 \\
\hline
3 & 4.52 & 7.48 \\
\hline
4 & 4.92 & 8.2 \\
\hline
5 & 5.48 & 9.0 \\
\hline
6 & 6.2 & 9.88 \\
\hline\hline
\end{tabular}
\end{center}
\end{minipage}
\begin{minipage}{0.26\linewidth}
\begin{center}
\begin{tabular}{|c|c|c|}
\hline\hline
$V$ &  $\hbar \omega_{1}$ & $ \hbar \omega_{2}$ \\
\hline\hline
0 & -- & -- \\
\hline
1 & 5.0 & 8.6 \\
\hline
2 & 5.64 & 11.72 \\
\hline
3 & 4.52 & 7.48 \\
\hline
4 & 4.92 & 8.2 \\
\hline\hline
\end{tabular}
\end{center}
\end{minipage}
\end{table}

\begin{figure}[!h]
\centerline{\includegraphics[width=0.425\columnwidth]{SiV4Mu4A0d1.eps} \quad
\includegraphics[width=0.425\columnwidth]{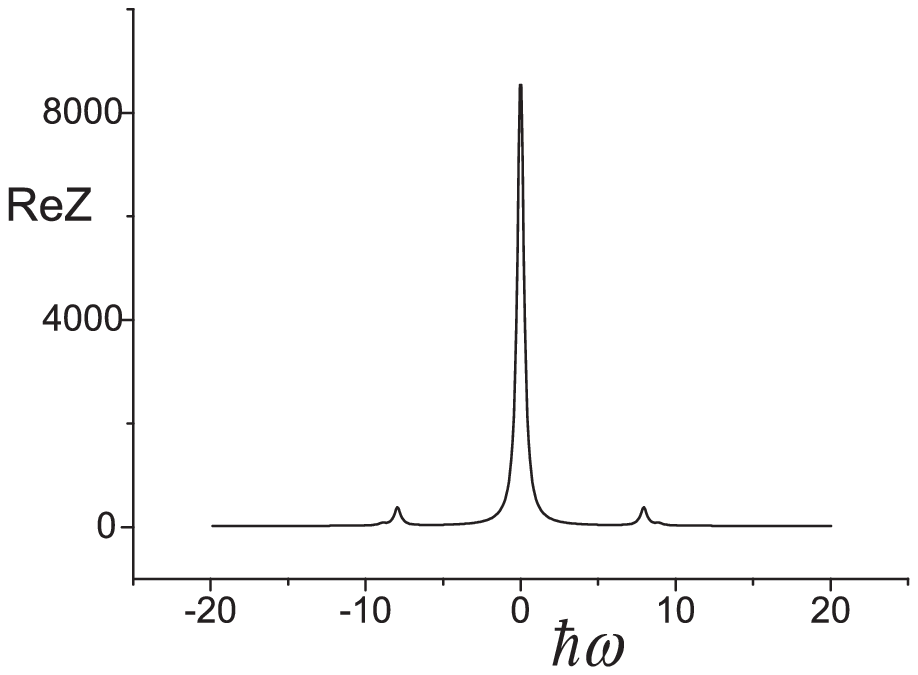}} \vspace{3mm}
\centerline{\includegraphics[width=0.425\columnwidth]{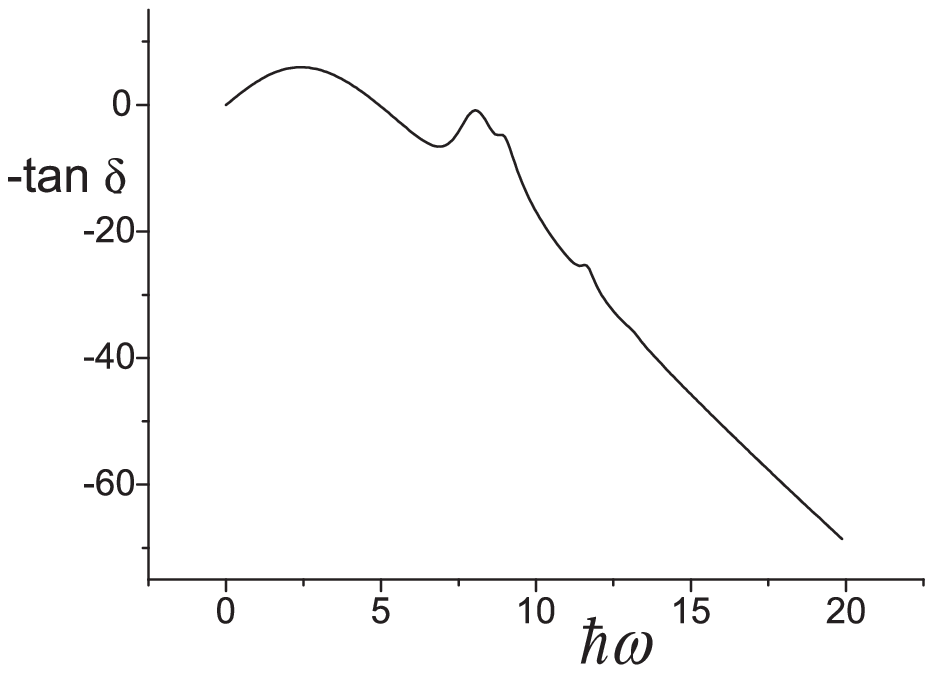}}\vspace{3mm}
\centerline{\includegraphics[width=0.425\columnwidth]{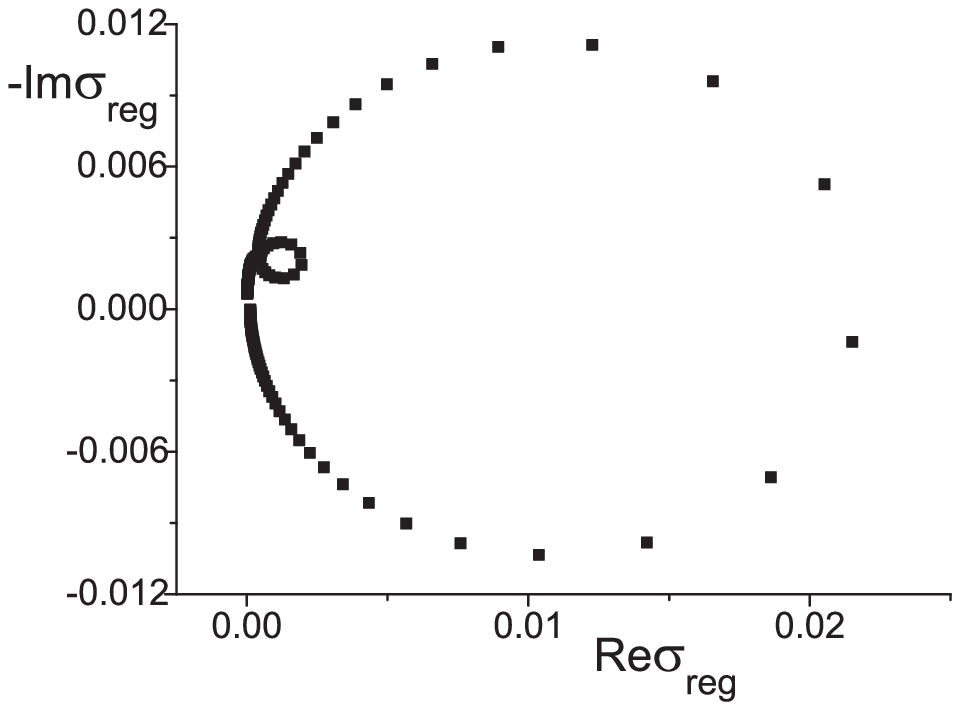} \quad
\includegraphics[width=0.425\columnwidth]{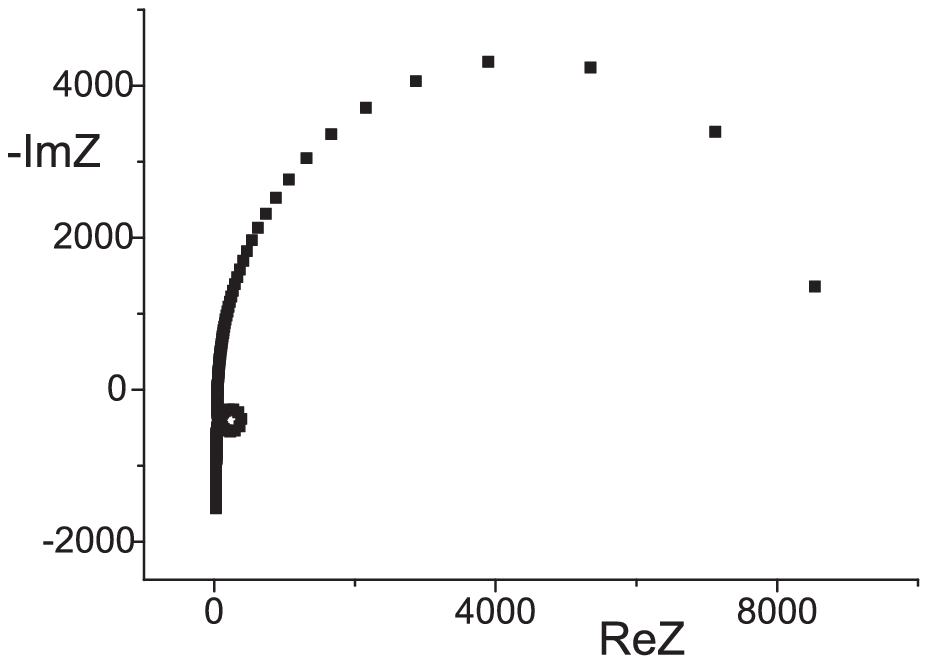}}
\caption{Frequency dependences of the
real parts of dynamic conductivity $\Re\sigma_{\text{reg}}$, impedance $\Re Z$ and the loss tangent $\tan\delta$;  Nyquist diagrams  for CDW phase ($\mu' = 0$);
$N = 10$, $V = 4$, $A = 0$, $t = 1$, $\Delta = 0.25$.} \label{fig3}
\end{figure}

Dynamic  conductivity of one-dimensional ionic chain is calculated
according to the formula (\ref{sigma}). The obtained frequency
dependence of the real part of the dynamic conductivity $\Re
\sigma_{\text{reg}}(\omega)$ of our one-dimensional model at $T = 0$ is
presented in figure~\ref{fig2}. Here, as an example, the  results
are shown for the case $V = 4$. Two maxima of conductivity (if we
consider only positive frequencies $\omega$) are obtained. The diagram in figure~\ref{fig1} shows that the  one-dimensional chain
at $\mu' = 0$ is a dielectric in the charge density wave (CDW) state.
In the absence of a modulating field ($A = 0$), two maxima of
conductivity are positioned at frequencies $\hbar \omega_{1} =
4.92$, $\hbar \omega_{2} = 8.12$. When the modulating field is
included, the  peak values increase becoming the largest at the
field $A \approx 1$. At the further growth of a modulating field, the
conductivity monotonously decreases and its two maxima shift to
the region of higher frequencies (when $A > 2$, one of the peaks
practically disappears). In the superfluid phase, the peaks of
conductivity are located in the nearly the same frequency region
but are slightly displaced. However, their weights are
significantly changed. While in CDW phase there is a higher peak
at lower frequencies, in the SF phase, on the contrary, the
opposite picture is observed. In addition, what is more important,
the heights of the peaks are significantly less in the SF phase
(up to two orders at $A = 0$). When the modulating field is
present in SF phase ($\mu' = -5$), the heights of the peaks significantly
increase, similarly to the case of CDW phase [at $A = 2$ we are
passing from SF phase to CDW phase (see figure~\ref{fig1}) and
obtain the conductivity peaks which are characteristic of
CDW phase]. Tables~\ref{tb3} and \ref{tb4} present the frequencies at
which the peaks of dynamic conductivity are located:
for $V = 4$ and $V = 0$, depending on the values of the modulating
field $A$, and for $A = 0$, depending on the values of the
interaction $V$ (table~\ref{tb5}). As can be seen from the tables,
the increase of interaction $V$ constant and the increase  of the
modulating field $A$ lead to the shift of the peaks of the dynamic
conductivity in the direction of higher frequencies. For example,
in the case $V = 0$ (table~\ref{tb3}), we have only one peak of
dynamic conductivity in CDW phase; a rough estimate shows the
linear dependence of the peak position on the modulating field [$\hbar
\omega_{1} (A) \approx 2A$].  In the case of SF phase, the
behaviour of the peaks is more complicated (see table~\ref{tb4}).
Here, the inclusion of a modulating field initially leads to a shift
of the peaks towards  lower frequencies. At $A = 2$ and $V = 4$,
we  come from the SF phase to CDW phase and the peak position
coincides with the relevant one for CDW phase (see table~\ref{tb3} and \ref{tb4}).
The shift of the peaks affected by interaction
$V$ is much weaker. For $A = 0$ in CDW phase, it is approximately
proportional to $V$ (see table~\ref{tb5}). In particular, $\hbar
\omega_{1}(V)$ is of the order of $V$, which is in accord with
the previous results for conducting phases of 1d models \cite{Fye}
and reflects the similar microscopic nature of the peak
appearance.

\begin{figure}[!h]
\centerline{\includegraphics[width=0.425\columnwidth]{SiV4Mu1A0d1.eps} \quad
\includegraphics[width=0.425\columnwidth]{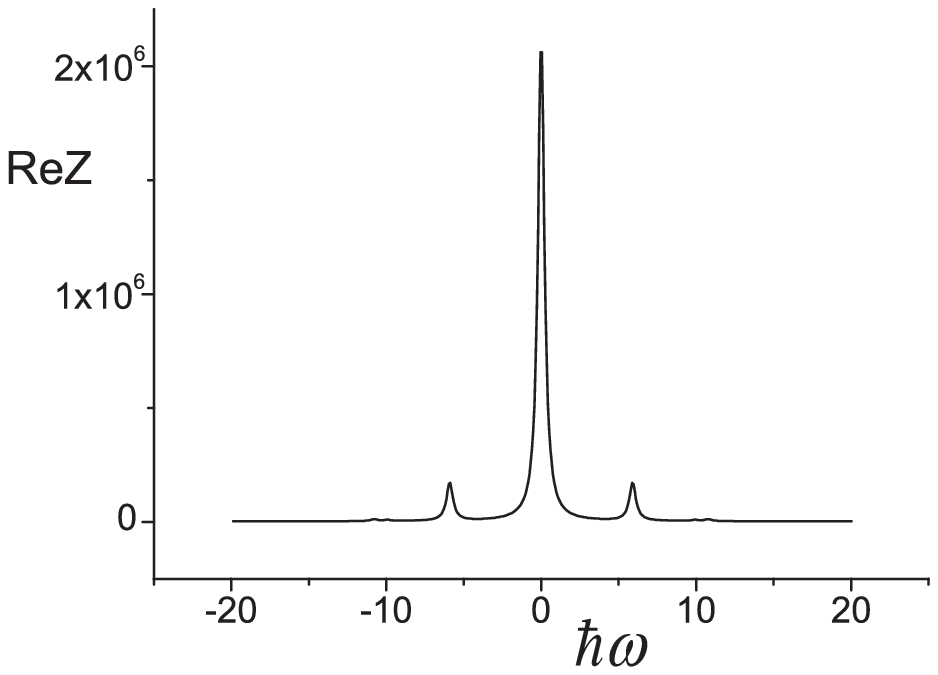}} \vspace{3mm}
\centerline{\includegraphics[width=0.425\columnwidth]{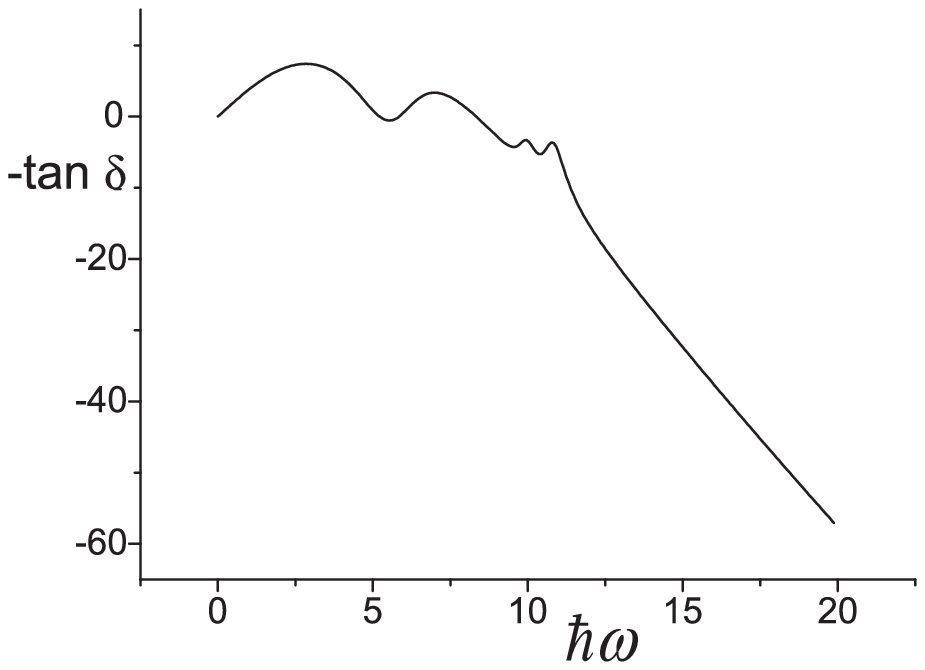}}\vspace{3mm}
\centerline{\includegraphics[width=0.425\columnwidth]{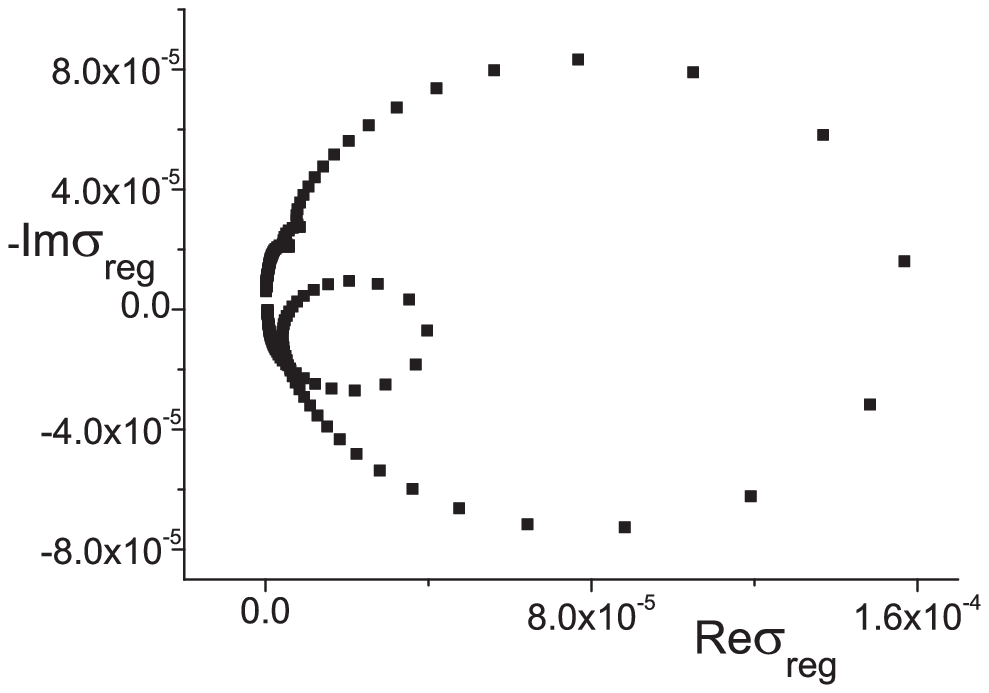} \quad
\includegraphics[width=0.425\columnwidth]{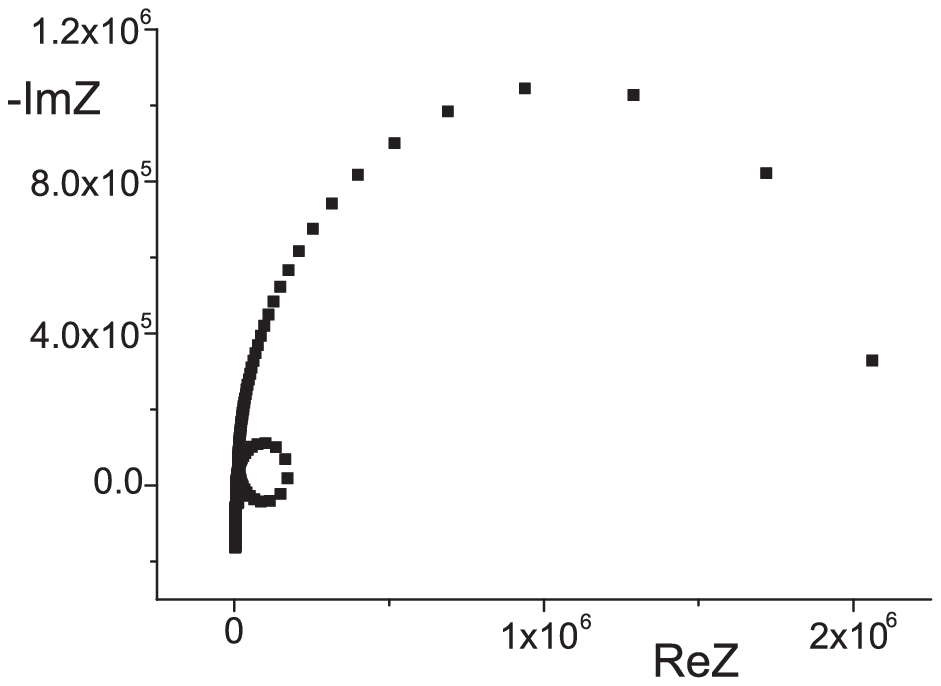}}
\caption{Frequency dependences of the
real parts of dynamic conductivity $\Re\sigma_{\text{reg}}$, impedance $\Re Z$ and the loss tangent $\tan\delta$;  Nyquist diagrams  for SF phase ($\mu' = -5$);
$N = 10$, $V = 4$, $A = 0$, $t = 1$, $\Delta = 0.25$.} \label{fig4}
\end{figure}

In figures~\ref{fig3}, \ref{fig4} we presented the calculated
frequency dependences of $\Re\sigma$, $\Re Z$ and $\tan\delta$, as well as
the Nyquist diagrams for impedance and conductivity in the case of
CDW and SF phases. As it follows from the figures, there exists a
direct correspondence between the peaks of conductivity and the ellipses
or semiellipses in Nyquist plots. In phenomenological description,
the existence of two ellipses can be interpreted as manifestation
of two types of collective vibrations (``oscillators'') in the
system, which exists due to the interaction between particles. In this connection it is
worth mentioning that in the absence of
interaction, there is no dynamical part of conductivity in one-dimensional
case.

\newpage
\section{Conclusions}

In this paper, using the exact diagonalization method we calculate the
energy spectrum and eigenfunctions of the finite one-dimensional
model of the ionic conductor with periodic boundary conditions. On
the basis of its eigen-states and with the help of Kubo formula we
calculate the dynamic conductivity  at  different values of the
interaction constant between particles and the strength of
the modulating field. The consideration is performed within the
hard-core boson approach (that corresponds to the Pauli
statistics). Calculations are  made for the case $T = 0$.

Based on the results of the energy spectrum calculations
\cite{Stetsiv}, we expect that at  the transition from CDW-type
phase to the superfluid phase, the peaks of dynamic conductivity  of
the ion chain will shift to the region of lower frequencies.
Indeed, in SF phase there are  transitions at lower energies
than in the CDW phase, but the selection rules for matrices
$A^{i}_{pq}$ practically eliminate their contributions to
conductivity. Similar results were also obtained by other authors
(see for example \cite{Kuhner}).

Frequency dependence of dynamic conductivity $\Re
\sigma_{\text{reg}}(\omega)$ remains qualitatively unchanged after the
transition from CDW phase to SF phase.  Only the heights of the maxima of conductivity are changed
significantly. In
particular, the heights of the peaks are in SF phase of the one order
(at $A = 1$), or even of the two orders (at $A = 0$) less than those in
CDW phase. The peaks of the dynamic conductivity in CDW phase
shift in the direction of higher frequencies when the  values of
interaction between the ions $V$ and the modulating field $A$ increase.
The effect of the modulating field is here much larger. The peak
shift is directly proportional to the field $A$ at $V = 0$. As
can be seen from the Kubo formula, Drude weight [static
conductivity $\Re \sigma(\omega = 0)$] is non-zero if there are
degenerate states. At $T = 0$, this is possible when  the ground
state is degenerate. We analyzed the energy spectrum obtained in
different phases of the finite one-dimensional model of the ion
conductor. In all cases, the ground state is nondegenerate (which
can be due to finite size of the ion chain), and the  Drude weight
calculations based on the Kubo formula give the $D = 0$ result.
Therefore, to calculate the Drude weight we use an alternative way
(based on the sum rule) that connects Drude weight with the
frequency integrated  regular part of conductivity and the average
kinetic energy of the ions. In this case, our calculations of $D$
for the half-filled CDW phase still give a non-zero result
(which is caused by a finite size of a system), but the obtained
value of $D$ is significantly smaller than the one for SF phase [here,
the two terms in (\ref{DDD}) nearly compensate each other]. A
significant reduction of the dynamic conductivity in SF phase
(compared with CDW phase) leads to a considerable reduction of the
second term of Drude weight [$ \int\Re \sigma_{\text{reg}}(\omega)
\rd\omega $] in SF phase  and, therefore, the static conductivity  in
SF phase is mainly determined by the kinetic energy of ions
$\frac{1}{N} \langle - \hat T \rangle $.

Besides the real part of conductivity, we have also calculated the
impedance $Z(\omega) = 1/\sigma(\omega)$ and loss tangent
$\tan\delta = \Im Z(\omega)/\Re Z(\omega)$. Frequency behaviour
of such quantities is important from the point of view of
experimental study. In order to interpret the results of the measurements,
the Nyquist diagrams (the $\Im Z$ versus $\Re Z$ plots) are usually
used. In particular, an important feature of our diagrams is a
significant difference in the frequency range of Nyquist plots for
CDW and SF phases (up to two orders, as is seen from the presented
figures). In practice, it can be used for identifying the state of
the ion conducting system.

\ukrainianpart

\title{Динамічна провідність одновимірних іонних провідників.
 Імпеданс, діаграми Найквіста}

\author{І.В. Стасюк, Р.Я. Стеців}
\address{Інститут фізики конденсованих систем НАН України,
вул. Свєнціцького, 1, 79011 Львів, Україна}

 \makeukrtitle

 \begin{abstract}
 \tolerance=3000%
 Досліджено залежність динамічної провідності одновимірного іонного
 провідника в залежності від величини взаємодії між частинками і
 величини модулюючого поля. Розгляд базується на гратковій моделі
 жорстких бозонів. Розрахунки проведено методом точної
 діагоналізації для скінченного одновимірного кластера. Вивчається
 частотна залежність динамічної провідності і поведінка її
 статичної складової (внеску Друде) в зарядовпорядкованій фазі
 (CDW) і в фазі типу суперфлюїду (SF). Розраховано частотну
 дисперсію імпедансу і тангенса втрат; побудовано діаграми
 Найквіста.
 \keywords іонний провідник, модель жорстких бозонів, внесок Друде,
 динамічна провідність, діаграми Найквіста

 \end{abstract}
\end{document}